# Terahertz Induced Photoconductivity of 2D Electron System in HEMT at Low Magnetic Field

Andrey Chebotarev and Galina Chebotareva

*OpthUS, P.O. Box 20042, Stanford, CA 94309, USA*

**Abstract.** A few results of our study of two-dimensional electron system (2DES) in low magnetic fields in GaAs/GaAlAs heterostructures by cyclotron resonance (CR) and photoconductivity techniques are presented. We have first discovered "CR-vanishing effect" in 2DES as well-defined crevasse on CR line in low magnetic fields, when Hall resistance is not quantized. "CR-vanishing effect" indicates vanishing longitudinal resistance & conductivity in these magnetic fields. Observed "CR-vanishing effect" demonstrates new correlated state of electrons in 2DES.

## INTRODUCTION

Our study of 2D electron systems in low magnetic field in GaAs/AlGaAs HEMT heterostructures, having fundamental physics interest, is directed to development of new kind of THz devices. First results have been obtained by us since 1992 when we started to study Cyclotron Resonance (CR) of 2DES with high mobility at low magnetic fields. Interest of scientific society to behavior and properties of 2DES in GaAs/AlGaAs at low magnetic fields has been dramatically increased lately due to "microwaves induced vanishing resistance effect" discussed in literature [3].

## RESULTS AND DISCUSSION

We have measured CR and Shubnikov-de-Haas (SdH) oscillations induced by THz radiation in samples of GaAs/GaAlAs heterostructures (MBE), with electron density $3-5 \times 10^{11} cm^{-2}$, mobility $10^5$ and $10^6$ $cm^2 V^{-1} s^{-1}$ at low magnetic fields around 0.3 T, at temperature $4,2K^0$ and frequencies 0.13 – 0.15 THz by photoconductivity technique (Fig.1-3). We have observed the vanishing of photocurrent & photovoltages as well-defined crevasse on CR line before CR maximum position in 2DES with mobility $10^6$ $cm^2 V^{-1} s^{-1}$ (Fig1). We have observed that effect both in photovoltaic and photoconductivity measurements. The effect keeps value under attenuation THz power on 10 dB (Fig. 2), but does not appear in samples with lower mobility $10^5$ $cm^2 V^{-1} s^{-1}$ (Fig.3). "CR-vanishing effect" indicates vanishing longitudinal resistance & conductivity (going to zero) in the limited regions of low magnetic fields. Hall resistance ($R_H$) has not been quantized at these low magnetic fields vs. $R_H$ in experiments [4] at high magnetic fields. Observed "CR-vanishing effect" demonstrates new correlated state of electrons in 2DES.

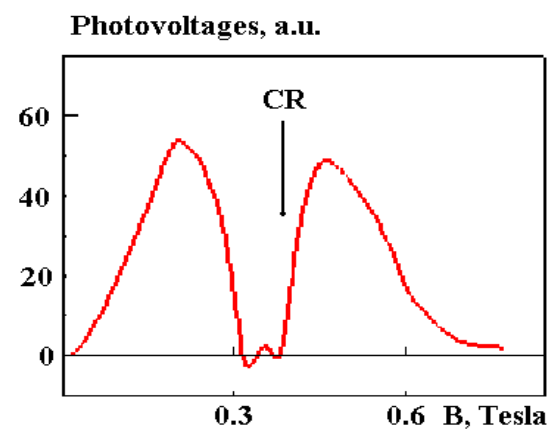

**FIGURE 1.** Magnetospectrum CR at 0.13 THz for 2DES with mobility $10^6$ $cm^2 V^{-1} s^{-1}$

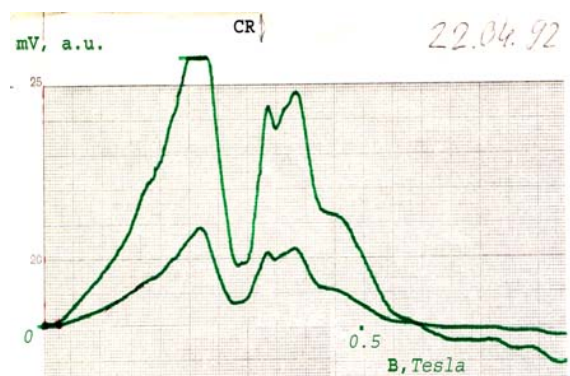

**FIGURE 2.** Magnetospectrum CR at 0.14 THz for 2DES with mobility $10^6$ $cm^2 V^{-1} s^{-1}$. Lover curve was been measured under attenuation THz power on 10 dB. The both curves are similar in low fields including CR-vanishing region.

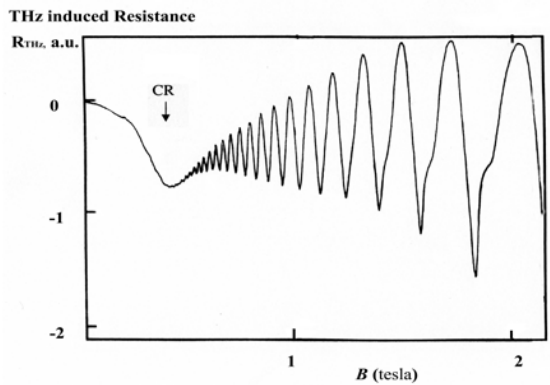

**FIGURE 3**. THz induced negative resistance in 2DES. CR and SdH oscillations caused by incident THz electromagnetic waves in 2DES with lower mobility ($10^5$ cm$^2$V$^{-1}$s$^{-1}$) at temperature $4,2K^0$ and F=0.13 THz. The magnetospectrum demonstrates very sharp line of CR without peculiarities except regular SdH oscillations that resolved up to maximum of CR-line.

Our research demonstrates that THz /Microwave radiation causes both photocurrent and photovoltages in 2DES of GaAs/AlGaAs heterostructures in low magnetic fields. Therefore, "microwaves induced vanishing resistance effect" (discussed now in literature [3]) could be dependent on a value and sign of photocurrent induced by the same radiation. Our results indicate that the photovoltaic response is connected with the sample surface area instead of edge effects [3, 5]. We have observed CR-vanishing effect both in photovoltaic and photoconductivity measurements.

## CONCLUSION

Our investigations of GaAs/AlGaAs heterostructures by cyclotron resonance and photoconductivity techniques have shown that the magnetospectra have the complex structure that depends on 2D-electrons' mobility & concentration, frequency and polarization of the incident radiation. We have demonstrated that THz/Microwave radiation induces both photocurrent and photovoltages in 2DES of GaAs/AlGaAs heterostructures. "CR-vanishing effect" (CRV) in 2DES at low magnetic fields, when the Hall resistance is not quantized, have been discovered by us. Observed "CR-vanishing effect" demonstrates new correlated state of electrons in 2DES. This CRV state is not depended on incident THz power.

## ACKNOWLEDGMENTS

Authors are thankful to professors A.M.Dykhne, V.I Gavrilenko, E. Gornik, K. von Klitzing, R.A.Stradling, and V.I.Ryzhii for useful discussions and interest to our research in different years.

**AUTHORS at ICPS-27, 2004**

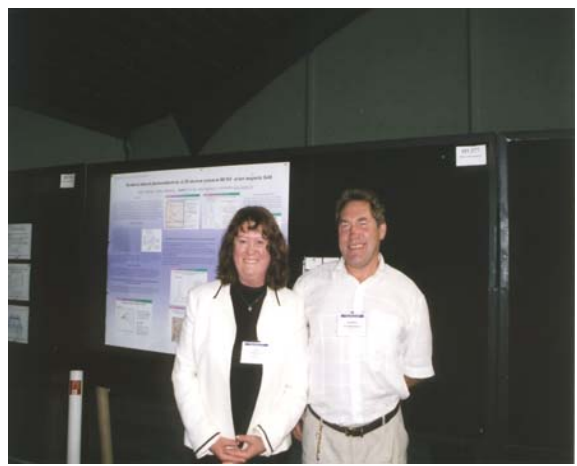